

\magnification=\magstep1

\def\D{\font\x=cmsy10\hbox{\x\char'104}}

\def\lb{\lbrack} \def\rb{\rbrack}

\def\Buildrel#1\over#2{\mathrel{\mathop{\kern 0pt #1}\limits_
{#2}}}

\def\truepageno{\footline={\hss\tenrm\folio\hss}}
\def\sqr#1#2{{\vcenter{\vbox{\hrule height.#2pt
      \hbox{\vrule width.#2pt height#1pt \kern#1pt
          \vrule width.#2pt}
        \hrule height.#2pt}}}}

\def\lapl{\mathchoice\sqr74\sqr74\sqr74\sqr74}

\def\\{\hfil\penalty-10000}
\nopagenumbers

\line{\hfill HD-THEP-92-28}
\vskip2cm
\centerline{\bf
NONPERTURBATIVE CONDITIONS FOR LOCAL WEYL INVARIANCE}
\centerline{\bf ON A CURVED WORLD SHEET}
\vskip2cm
\centerline{Jens Schnittger\footnote*{supported by DFG; \quad present address:
Ecole Normale Sup\'erieure, Laboratoire de
Physique Th\'eorique, 24 rue Lhomond,
F-75231 Paris Cedex 05} and
Ulrich Ellwanger\footnote{**}{supported by a DFG Heisenberg fellowship}}
\vskip.5cm
\centerline{Institut f\"ur Theoretische Physik}
\centerline{Universit\"at Heidelberg}
\centerline{Philosophenweg 16, D-6900 Heidelberg}
\vfill
\noindent{\bf ABSTRACT:}
We investigate Weyl anomalies on a curved world sheet to second order in
a  weak field expansion. Using a local version of the exact
renormalization  group equations, we obtain nonperturbative results for
the tachyon/graviton/dilaton system. We discuss the elimination of
redundant operators, which play a crucial role for the emergence of
target  space covariance. Performing the operator product expansion on a
curved  world sheet allows us to obtain the nonperturbative
contributions to the dilaton $\beta$ function. We find the $\beta$
functions,  after suitable  field redefinitions, to be related to a
target space effective action through a $\kappa$ function involving
derivatives.  Also we can establish a nonperturbative Curci-Paffuti
relation including the  tachyon $\beta$ function.
\eject
\truepageno\pageno=2
\medskip
1.) The intimate relation between critical string theory and Weyl invariant two
dimensional field theories has motivated numerous investigations of their Weyl
anomalies or renormalization group (RG) equations. The RG equations can be
interpreted as string equations of motion, and their fixed points determine
consistent string backgrounds \lb 1-4\rb.
\smallskip
Most studies were restricted to massless string modes, i. e.  renormalizable
$\sigma$-models, where the Weyl anomaly is characterized by the  standard
$\beta$ functions of a renormalizable 2d field theory. The computations  were
performed in a loopwise expansion employing the background field method,  which
keeps general coordinate invariance in the target space manifest. The  $\beta$
function of the dilaton plays a special role: It determines the central charge
of the  theory, and thus should be constant. This can indeed be shown to
be the case, provided
the  $\beta$ functions corresponding to all other massless modes vanish \lb
2,5,6\rb.
\smallskip
However, apart from being unable to describe massive modes, the loopwise
expansion has a problem of principle: It misses the contributions to the  Weyl
anomaly from nonperturbative infinities, which are not visible to any  finite
order in $\alpha'$ \lb 7\rb. These can be  obtained from a weak field
expansion, where the string partition function -  including a "background" of
vertex operators - is expanded in powers of  these vertex operators \lb
7-20\rb. The corresponding contributions to the  Weyl anomalies can be
summarized in the exact renormalization group  (ERG) equations \lb 21-27\rb,
which are not only nonperturbative in $\alpha'$ or the  number of loops on the
world sheet, but also exact in the number of vertex  operators. (The weak field
expansion is recovered by solving the exact      renormalization group
equations
iteratively, see below.)
\smallskip
Most results in the framework of the weak field expansion obtained up to now
are based on the investigation of (local) Weyl invariance on a flat world sheet
up to second order in the field strengths. The flat world sheet does not allow,
however, to obtain the analog of the dilaton $\beta$ function in any
straightforward manner. On the other hand, apart from containing the
information about the central charge c the dilaton  $\beta$ function is
required in order to discuss the proper  relations between the $\beta$
functions and a covariant action \lb 1,2,6\rb.  In \lb14\rb\  contributions to
the dilaton $\beta$ function in a weak field  expansion have been considered
for the massless modes of the superstring,
and  in \lb  20\rb \  in a perturbative
expansion to low orders in $\alpha'$. Therefore  the relation between
nonperturbatively derived $\beta$  functions including massive modes and a
covariant  action is still an open problem.
There are arguments
by Tseytlin \lb 28\rb \   against
the existence of such an action when all modes are included.
On the other hand, as pointed out in \lb 29\rb, a covariant action for
the graviton and tachyon alone may well exist, in spite of nonperturbative
terms in the $\beta$ functions  whose covariant form is rather non-obvious.
Irrespective of these difficulties, covariant actions and  $\beta$ functions
have been used recently
in studies of stringy  black holes including massive, i.e. tachyonic, matter
\lb
30-32\rb.
\smallskip
Recently we have  generalized the ERG equations to a curved world sheet by
formulating the  condition of local scale invariance as a condition of
independence of the  partition function on the Liouville mode $\sigma(z)$ \lb
33\rb. Here the inclusion  of $R^{(2)}$ dependent operators is straightforward,
and the central charge  appears correctly as the $\beta$ function associated
with the operator  $R^{(2)}$.
\smallskip
In this paper we will present a systematic analysis of iterative solutions  to
these equations. Since we do not make any gauge choice, we are able to address
the question of target space covariance. Here the redundant operators  (which
can be eliminated by  using the world sheet equations of motion) turn  out to
play a crucial role and have to be treated carefully. To the level we have
checked we find manifest target space covariance,
 at least to lowest order in spacetime momenta,
if the contributions of
redundant operators to the $\beta$ functions  are eliminated. If, on the other
hand, they  are required to vanish separately, this amounts to a gauge fixing
condition,  which corresponds to the harmonic gauge at the linearized level.
\smallskip
For the first time, we are able to  investigate the analog of the Curci-Paffuti
relation including the tachyon  explicitly. Such a relation is indeed seen to
hold, but its form is  considerably more complicated
and in fact suggests that general covariance may be  realized in a
nontrivially deformed fashion at the
nonperturbative level.
 Our results allow to
discuss the relation  between the $\beta$ functions and a target space
effective action including  the tachyon and the dilaton. We find the need to
incorporate a $\kappa$ function  into such a relation (cf. eq. (4.7) below),
where the $\kappa$ function contains  space-time derivatives.
\smallskip
In the next section we present the underlying formalism of our approach,  the
ERG equations and their solution within a weak field expansion.  In addition we
will discuss technical issues such as the preservation of world sheet
covariance through the process of regularisation and in the context of  the
operator product expansion on a curved world sheet. Furthermore, the proper
treatment of redundant operators is discussed in some detail.
 In section 3 we present our results to  first and second order in the weak
field expansion. Section 4 is devoted to  a discussion of the $\beta$
functions, their relation to a target space  effective action (via a $\kappa$
function) and the analog of a Curci-Paffuti  relation. A summary, conclusions
and an outlook are given in section 5.
\bigskip

2.) Let us start with the presentation of the local ERG equations, which have
been derived in \lb 33\rb\ from the condition of Weyl invariance of the matter
part of the Polyakov path integral $\hat Z$, with $\hat Z$ given by

$$\hat Z\lbrack\sigma\rbrack=e^{-26S_L\lbrack\sigma\rbrack}\int
\D Xe^{-(S_0+S_{int})}.\eqno(2.1)$$
Here we have d bosonic fields $X^\mu$ with a free action

$$S_0={1\over{8\pi}}\int d^2z\partial_z X^\mu\partial_{\bar
z}X^\mu\eqno(2.2)$$ and an arbitrary interaction $S_{int}$.
$S_L\lb\sigma\rb$ satisfies

 $${{\delta S_L\lbrack\sigma\rbrack}\over{\delta\sigma(z)}}=
{1\over{48\pi}}\cdot R(z)\eqno(2.3)$$
and originates from gauge fixing the world sheet metric $g_{\alpha\beta}$ into
the conformal gauge $g_{\alpha\beta}=\delta_{\alpha\beta}e^\sigma$. The
condition of $\sigma$-independence of $\hat Z$ translates into the following
condition on $S_{int}$:
  $$\eqalign{&{{d-26}\over{48\pi}}R(z)-{{\delta S_{int}}\over{\delta\sigma(z)}}
+{1\over2}\int d^2z_1\sqrt{g(z_1)}\int d^2z_2\sqrt{g(z_2)}
{{\delta G_{reg}(z_1,z_2)}\over{\delta\sigma(z)}}\cr
&\cdot\left({{\delta S_{int}}\over{\delta X^\mu(z_1)}}{{\delta S_{int}}
\over{\delta X^\mu(z_2)}}-{{\delta^2S_{int}}\over
{\delta X^\mu(z_1)\delta X^\mu
(z_2)}}\right)+\cr&+\int d^2z'\sqrt{g(z')}\left({{\delta F_\mu\lbrack
X, z,z'\rbrack}
\over{\delta X^\mu(z')}}-F_\mu \lbrack X,z,z'\rbrack
{{\delta S}\over{\delta
X^\mu(z')}}\right) = 0\cr}\eqno(2.4)$$
where $G_{reg}$ denotes the regularized (and hence $\sigma$-dependent)  bosonic
propagator and $S = S_0 + S_{int}$.
(The precise form of the regularization is still arbitrary at  this
stage.) The arbitrary functionals $F_\mu\lb X,z,z' \rb$ originate  from the
freedom to add total derivatives under the path integral. They can be related
to certain redefinitions of the bosonic fields $X^{\mu}$ and generate the
Schwinger-Dyson equations allowing to eliminate redundant operators.
\smallskip
To proceed, we assume that $S_{int}$  can be expanded
into a complete set of local
operators,
 $$\eqalign{&S_{int}=\sum_n \int d^2 z \sqrt{g(z)} G^n (X) P_n
(\nabla X(z),\sigma(z))\cr
&=\sum_n\int d^D p \tilde G^n(p)\int d^2z\sqrt{g(z)}V_n
(X(z),\sigma(z),p)}\eqno(2.5)$$
where
 $$V_n(X(z),\sigma(z),p)=e^{ipX(z)}P_n(\nabla X(z),\sigma(z)),
\eqno(2.6)$$
Here $P_n$ is a polynomial in covariant derivatives $\nabla$ of
$X^\mu(z)$ and may have additional non-trivial dependence on
$\sigma(z)$ via the two-dimensional curvature $R$ such that
$P_n$ is a scalar and $S_{int}$ respects reparametrization
invariance. In turn the l.h. side of eq.(2.4) can be expanded into the same
complete set of local operators, and the coefficient of each operator
will be called the $\beta$ function associated with the corresponding
operator subsequently.
Also, we shall assume in the following that appropriate powers of the
cutoff have been factored out of the couplings $G^n$ according to the
Wilson-Wegner prescription, so that the $G^n$ become dimensionless.
 From dimensional considerations, it is clear that these cutoff powers
can be absorbed by the metric dependence of eq.(2.5), i.e. only the combination
$\epsilon^2 e^{-\sigma (z)}$ will then appear in (2.5) (cf. \lb 33,54\rb).
To save notation, we therefore choose to display only the metric-dependence
in the following, the presence of the cutoff being understood implicitly.
\smallskip
In order to solve the ERG equations in a weak field expansion (or iteratively)
we have to introduce a small expansion parameter $\lambda$ and to make an
ansatz for the order of $\lambda$ of the background string modes $G^n$.
As background string modes we will take into account the dilaton $\Phi(X)$,
the graviton $H_{\mu\nu}(X)$ and the tachyon $T(X)$ so that $S_{int}$ reads
$$S_{int} = {1\over{8\pi}} \int d^2z \lb H_{\mu\nu}(X) \partial_z
X^\mu\partial_{\bar z}X^\nu + \lapl_z \sigma(z)
\Phi(X) + e^{\sigma}T(X) + ...\rb
 \eqno(2.7)$$
where the dots stand for an infinite set of other massive modes $M^n(X)$.
Now we will make the ansatz
$$\eqalign{&T(X),\Phi(X), H_{\mu\nu}(X) \sim O(\lambda^1),\cr
&M^n \sim O(\lambda^2) }\eqno(2.8)$$
for all other massive modes $M^n$. (In {\lb33\rb}, we
have already shown how to treat a dilaton of the form $\Phi(X) = Q_\mu X^\mu$
exactly.)
Below we will compute the $\beta$ functions to $O(\lambda^1)$ and
$O(\lambda^2)$  associated with the following four operators:
$$\eqalign{a) &\sim T(X) \cr
b)&\sim \partial X^\mu\partial X^\nu H_{\mu\nu}(X) \cr
c)&\sim \lapl_z X^\mu V_\mu(X) \cr
d)&\sim \lapl_z \sigma(z) \Phi(X).} \eqno(2.9)$$
(The $\beta$ functions associated with the other massive modes start to
$O(\lambda^2)$  only; their vanishing determines the necessarily non-vanishing
background $M^n$  in terms of $T, H_{\mu\nu} $ and $\Phi$. The background $M^n$
itself does not  appear in the four $\beta$ functions considered above to
$O(\lambda^2)$, cf. below.) Note that the conditions for local Weyl invariance
do not
allow for the addition of total (world sheet) derivatives  to these equations,
i.e. the coefficients of the operators b) and c)  have to vanish independently.
The operator $\lapl_z X^\mu$, however,  can be eliminated using the equations
of
motion of the field $X^\mu$ or,  alternatively, by choosing the functionals
$F^\mu \lb X,z,z' \rb $ in (2.4) such that the coefficient of $\lapl X^\mu$ in
eq. (2.4) vanishes identically (see below).
\smallskip
In the process of the evaluation of the Weyl invariance conditions, especially
of the quadratic term, it will be important to keep 2d reparametrization
invariance  manifest. On the other hand, the usual propagator    $G(z_1,z_2) =
-ln|z_1 - z_2|^2$ is not fully reparametrization invariant,  as is easily seen
by considering $SL_2({ C})$ - transformations of  $z_1, z_2$ (the only globally
defined coordinate transformations on the sphere which  respect the conformal
gauge). The same will of course be true for the regularized  version
$G_{reg}(z_1, z_2)$. The covariant propagator is of the form  $$G^{cov}(z_1,
z_2) = G(z_1, z_2) + f(z_1) + f(z_2) \eqno(2.10)$$ where $f(z)$ depends on the
definition of the zero mode of $X^\mu$. In the  following we shall work with
$$f(z) = - {1\over 2} \sigma(z), \eqno(2.11)$$
a choice corresponding (up to an irrelevant constant piece) to the Arakelov
propagator \lb 34\rb. $f(z)$ will eventually cancel in expectation values due
to the  zero mode integration \lb 35\rb, but at intermediate stages covariance
will be manifest only  when it is taken into account. Concerning the variations
with respect to  $\sigma(z)$, $\delta G_{reg}(z_1, z_2) / \delta \sigma (z)$ is
covariant by  itself, since this is also true for $\delta
(f(z_1)+f(z_2))/\delta \sigma (z)$.  Hence we can replace $\delta
G_{reg}^{cov}(z_1, z_2) / \delta \sigma (z)$ by  $\delta G_{reg}(z_1, z_2) /
\delta \sigma (z)$ in eq. (2.4) in the terms  linear and quadratic in $S_{int}$
without loosing covariance. (Since the above- mentioned cancellation of $f(z)$
holds for propagators with Weyl derivatives  separately, we can keep $f(z)$ in
all other contractions.)
 \smallskip
Now we turn to the treatment of the term in eq. (2.4) quadratic in $S_{int}$.
As outlined in \lb 24\rb (see also \lb33\rb), we first have to normal order
and  then to Taylor expand  this term into a complete set of local operators.
After the decomposition (2.5)  of $S_{int}$ the term is quadratic in the
operators $V_n(z, \sigma, p)$, and  the result of their normal ordering can be
written as  $$V_m(z_1, \sigma, p_1) V_n(z_2, \sigma, p_2) = C_{m
n}(z_1,z_2,\sigma,p_1,p_2)  N(z_1,z_2,\sigma,p_1,p_2) \eqno(2.12)$$ Here
$N(z_1,z_2,\sigma,p_1,p_2)$ denotes the normal ordered product of two operators
located at $z_1$ and $z_2$, and $C_{m n}(z_1,z_2,\sigma,p_1,p_2)$  is a
functional of the regularized Green functions $G^{cov}_{reg}$ and its
derivatives. $N(z_1,z_2,\sigma,p_1,p_2)$ can now be expanded
in terms of operators located at $z$ using the covariant Taylor expansion  (or
geodesic expansion):
$${N(z_1,z_2,\sigma,p_1,p_2) = \sum_k \alpha_k(z,z_1,z_2,\sigma,p_1,p_2) \cdot
V_k(z,\sigma,p_1+p_2)} \eqno(2.13)$$
Finally we obtain
$$V_m(z_1,\sigma,p_1) V_n(z_2,\sigma,p_2)=\sum_kC^k_{m
n}(z,z_1,z_2,\sigma,p_1,p_2) V_k(z,\sigma,p_1+p_2) \eqno(2.14)$$
with $C^k_{m n}=C_{m n}\cdot \alpha_k$. Thus, after inserting the decomposition
(2.5)  into the quadratic term in (2.4), the latter can be expressed as
$$\eqalign{& \int d^2z_1\sqrt{g(z_1)}\int d^2z_2\sqrt{g(z_2)}
{{\delta G_{reg}(z_1,z_2)}\over{\delta\sigma(z)}}
\cdot{{\delta S_{int}}\over{\delta X^\mu(z_1)}}{{\delta S_{int}}
\over{\delta X^\mu(z_2)}} \cr
&= \int d^D p_1 \int d^D p_2\tilde G^n(p_1) \tilde G^m(p_2)
\sum_k \tilde C^k_{m n}(z, \sigma,
 p_1, p_2) V^k(z, \sigma, p_1 +p_2) }\eqno(2.15)$$
with
$$\tilde C^k_{m n}(z, \sigma, p_1, p_2) = \int d^2z_1\sqrt{g(z_1)}
\int d^2z_2\sqrt{g(z_2)} {{\delta^{s.d.}}\over{\delta\sigma(z)}}
C^k_{m n}(z,z_1,z_2,\sigma,p_1,p_2), \eqno(2.16)$$
where ${{\delta^{s.d.}}\over{\delta\sigma(z)}}$ acts only on the $\sigma$
dependence  induced by the regularization of the short-distance singularity of
$G(z_1,z_2)$.  In a next step we can expand $\tilde C^k_{m n}(z, \sigma, p_1,
p_2)$ into local  covariant functionals of the world sheet metric. Thereafter
the right hand side of  eq. (2.15) becomes
$$\eqalign{& \int d^D p_1 \int d^D p_2\tilde G^n(p_1) \tilde G^m(p_2)  \sum_k
\hat V^k(z,\sigma, p_1 +p_2) \cr  & (\tilde C^{k (1)}_{m n}(p_1, p_2) + \tilde
C^{k (2)}_{m n}(p_1, p_2) R^{(2)}(z) +\cr  & \tilde C^{k (3)}_{m n}(p_1, p_2)
(R^{(2)}(z))^2 +  \tilde C^{k (4)}_{m n}(p_1, p_2) \lapl  R^{(2)}(z) + ...) .
}\eqno(2.17)$$
The set of operators $\lb \hat V_k \rb $ is, by definition, the subset of  $\lb
V_k \rb $ not containing any curvature factors; these appear explicitly  in the
second factor  on the right hand side of (2.17) ordered according to the number
of derivatives.  The contribution, e.g., to the dilaton $\beta$ function is
given by the coefficient  $\tilde C^{k (2)}_{m n}(p_1, p_2)$ where $k$ is
chosen such that  $\hat V^k(z,\sigma, p_1 +p_2)$ is the tachyonic vertex
operator.  The coefficients $\tilde C^{k (i)}_{m n}(p_1, p_2)$ will be called
operator product coefficients (OPCs) subsequently.
They depend only on the momenta, but not on the metric or cutoff.
\smallskip
The form of the expansion (2.17) looks very natural; neverless there are some
subtleties involved. It is clear that the left hand side of (2.15) can always
be expressed in the form $\sum_k \tilde C^k_{m n}(z, \sigma,
 p_1, p_2) \hat V^k(z, \sigma, p_1 +p_2) $ with some functional $\tilde C^k_{m
n} (z, \sigma, p_1, p_2)$ of the 2d metric. What is not clear is that we can
expand   $\tilde C^k_{m n}$ into positive integer powers of the curvature and
its  derivatives localized at $z$, which would lead to (2.17). Considering the
specific example where $ V_m,  V_n $ and $\hat V_k$ are all tachyonic
operators and  using the regularization discussed below, we find for the
constant curvature  metric
$$e^{\sigma (z)} = R^4/(R^2 + |z|^2/4)^2\eqno(2.18)$$
(R is the radius of the sphere),
$$\tilde C^k_{T T}(z,\sigma, p_1, p_2) = 1-(1+{{4R^2}\over{\epsilon^2}})^
{1+p_1 p_2} \eqno(2.19)$$
The quantity $R^2$ on the right hand side of (2.18) can have two  different
origins: First, it could be the remnant of some globally defined
quantity as
the volume of the world sheet, and hence be due to the regularization  in the
infrared region. Second, it could represent a local  curvature dependent
operator as one of those of eq. (2.17), since in the present  case we have
$R^{(2)} = 2 R^{-2}$ from eq. (2.18). In the first case it would  be sensible
to discard the first term in eq. (2.19), which would correspond  to a
particular prescription for the computation of the $\beta$ functions
(disregarding the effects of nonlocal globally defined operators as the volume)
also employed in ref. \lb 13\rb.
\smallskip
Actually we can decide between the two alternatives by observing that on
general  grounds the OPCs $\tilde C^k_{T T}(z)$ should be finite at points $z$
with  $R^{(2)}(z) = 0$.
 If we would replace $2 R^{-2}$ by $R^{(2)}(z)$ in eq. (2.19), on the other
hand, $\tilde C^k_{T T}(z)$ would diverge at these points provided $1+p_1 p_2
>0$.  Hence, we are lead towards the first interpretation of $R^2$ in  eq.
(2.19). Performing thus the  limit $\epsilon^2 / R^2 \to 0$ in eq. (2.19) (in
the kinematic region where  $1+p_1 p_2 <0$ ) , we obtain the standard unit
result
for the OPC coefficient $\tilde C^T_{T T}$. Furthermore the coefficients
$\tilde C^{k (i)}_{T T}(p_1, p_2)$ with $i$ corresponding to operators
containing powers of $R^{(2)}(z)$ (but no derivatives thereof) vanish in this
particular  case, which is, however, not generic.
\smallskip
A slight generalization of these arguments allows us to adopt the following
strategy for the computation of $\tilde C^{k (1)}_{m n}(p_1, p_2)$ and  $\tilde
C^{k (2)}_{m n}(p_1, p_2)$: We use the constant curvature metric as  above and
expand the $integrand$ of $\tilde C^{k}_{m n}(p_1, p_2)$ around the  flat space
limit, i. e. in powers of $\epsilon^2 /R^2$. In this way we   automatically
discard the parts of $\tilde C^{k}_{m n}(p_1, p_2)$ which are  not analytic in
$R^2$ and which can have nothing to do with the curvature  dependent operators
of eq. (2.17). The remaining integrals are computed for  $p_1 p_2$ such that
they converge in the infrared and defined by analytic  continuation for other
values of  $p_1 p_2$.
In the example of $\tilde C^k_{T T}(z)$  considered above
this prescription leads, of course, again to the result that  the coefficient
$\tilde C^{k (2)}_{T T}(p_1, p_2) =  \tilde C^{\Phi}_{T T}(p_1, p_2)$ vanishes.
We do not get any   information about the the OPC coefficients into operators
involving derivatives  of the curvature since we are using a constant curvature
metric, but these will not   be relevant for our analysis of the $T,
H_{\mu\nu}, \Phi$ system.
\smallskip
Let us now discuss the elimination of the redundant operator $\lapl_z X^\mu$
through  an appropriate choice of the functional $F^\mu \lb X,z,z'\rb $ in eq.
(2.4).  $F$ will be determined order by order in the weak field expansion, i.
e.  we make the ansatz $F = F^{(1)} + F^{(2)} + ...$ where the upper index
denotes  the corresponding power in the parameter $\lambda$ of eq. (2.8). Let
us introduce  the abbreviation $V_\mu(z)$ for the contributions to the  $\beta$
function associated with the operator  $\lapl_z X^\mu$, which originate from
the
parts involving $S_{int}$ in eq. (2.4).
 At the moment, we will consider only the first order contribution  to $V_\mu$,
i. e. $V_\mu \equiv V_\mu^{(1)}$. There are in principle second order
contributions arising from the quadratic term in (2.4). These will appear in
the  normal ordered form $:\lapl X^\mu V_\mu^{(2)}:$ and hence would seem to be
of $O(\lambda ^3)$  by use of the equations of motion. However, there is a
slight subtlety involved here,  related to our use of the covariant propagator
$G^{cov}_{reg}(z_1, z_2)$, which  will be discussed below.
\smallskip
In order to cancel the contributions proportional to $\lapl_z X^\mu$
in eq. (2.4) to first order in $\lambda$, $F$ has to satisfy
$$\eqalign{& \int d^2z'\sqrt{g(z')}\left({{\delta F_\mu^{(1)}\lbrack
X, z,z'\rbrack}
\over{\delta X^\mu(z')}}-F_\mu^{(1)} \lbrack X,z,z'\rbrack
{{\delta S_0}\over{\delta
X^\mu(z')}}\right) = \cr
& -V_\mu(z) \lapl_z X^\mu(z) + . . .}\eqno(2.20)$$
where the dots denote terms independent of $\lapl X$. It is important to note
that $\delta S_0/\delta X^\mu$ is given by  $-\lapl^{reg} X^\mu/(4\pi)$ where
$\lapl^{reg}$ is the inverse of the $regularized$  propagator $G_{reg}$.
Accordingly we define $1\!{\rm l}_{reg}(z, z')$ by
$$1\!{\rm l}_{reg}(z, z') = \lapl_z G_{reg}(z, z'). \eqno(2.21)$$
With the ansatz
$$F^{(1)}(z, z')=-4\pi 1\!{\rm l}_{reg}(z, z') \hat F^{(1)}(z), \eqno(2.22)$$
we find that $\hat F^{(1)}$ should be given by
$$\hat F^{(1)}(z)  = V^\mu(z) \eqno(2.23)$$
With this form of $F$ the first term in eq. (2.20) becomes
$$ \int d^2z'\sqrt{g(z')} {{\delta F_\mu^{(1)} \lbrack X, z,z'\rbrack}
\over{\delta X^\mu(z')}} = -4\pi V_{\mu ,\mu}(z) 1\!{\rm l}_{reg}(z, z).
\eqno(2.23)$$
To second order in $\lambda$ we have to consider the dependence
on $\lapl _z X^\mu(z)$  of
$$\eqalign{& \int d^2z'\sqrt{g(z')}\lbrack{{\delta F_\mu^{(2)}\lbrack
X, z,z'\rbrack}
\over{\delta X^\mu(z')}}-F_\mu^{(2)} \lbrack X,z,z'\rbrack
{{\delta S_0}\over{\delta X^\mu(z')}}
-F_\mu^{(1)} \lbrack X,z,z'\rbrack
({{\delta S_{int}}\over{\delta X^\mu(z')}})^{(1)}\rbrack\cr }\eqno(2.24)$$
with, using (2.7),
$$({{\delta S_{int}}\over{\delta X^\mu(z')}})^{(1)}=
-{{H_{\mu\nu}(z')\lapl_{z'}
X^\nu(z')}\over{4\pi}}+{{W_\mu(z')}\over{4\pi}} \eqno(2.25)$$
where $W_\mu$ is independent of $\lapl X$. Accordingly the last term on the
left
hand side of eq. (2.24) can be written as
$$\eqalign{& \int d^2z'\sqrt{g(z')} F_\mu^{(1)} \lbrack X,z,z'\rbrack
({{\delta S_{int}}\over{\delta X^\mu(z')}})^{(1)} \cr
&= V_\mu\int d^2z'\sqrt{g(z')} 1\!{\rm l}_{reg}(z, z')
(H_{\mu\nu}(z')\lapl_{z'}X^\nu - W_\mu(z'))\cr
& = V_\mu\int d^2z'\sqrt{g(z')} (\tilde H_{\mu\nu}(z, z')\lapl_{z'}^{reg}
 X^\nu - W_\mu(z')) 1\!{\rm l}_{reg}(z, z')}\eqno(2.26)$$
where
$$\tilde H_{\mu\nu}(z, z') = \int d^2z''\sqrt{g(z'')} 1\!{\rm l}_{reg}(z, z'')
H_{\mu\nu}(z'') 1\!{\rm l}_{reg}(z'', z').\eqno(2.27)$$
Let us now choose
$$F^{(2)}_\nu(z, z') = 4\pi V_\mu(z)\tilde H_{\mu\nu}(z, z') \eqno(2.28)$$
For the first term on the left hand side of eq. (2.24) we then obtain
$$\eqalign{& \int d^2z'\sqrt{g(z')}{{\delta F_\mu^{(2)}\lbrack  X,
z,z'\rbrack}\over{\delta X^\mu(z')}} = 4\pi V_{\mu,\nu}(z)\tilde H_{\mu\nu}(z,
z) \cr &+4\pi V_\mu(z)\int d^2z'\sqrt{g(z')} 1\!{\rm l}_{reg}(z, z')
H_{\mu\nu,\nu}(z')  1\!{\rm l}_{reg}(z', z')}\eqno(2.29)$$
Altogether we thus find
$$\eqalign{& \int d^2z'\sqrt{g(z')}\left({{\delta F_\mu\lbrack  X, z,z'\rbrack}
\over{\delta X^\mu(z')}}-F_\mu \lbrack X,z,z'\rbrack {{\delta S}\over{\delta
X^\mu(z')}}\right) - V_\mu(z) \lapl_z X^\mu(z) = \cr &4\pi V_{\mu ,\mu}(z)
1\!{\rm l}_{reg}(z, z) - V_\mu(z) \int d^2z'\sqrt{g(z')}  1\!{\rm l}_{reg}(z,
z') W_\mu(z') - 4\pi V_{\mu,\nu}(z)\tilde H_{\mu\nu}(z, z) \cr &-4\pi
V_\mu(z)\int d^2z'\sqrt{g(z')} 1\!{\rm l}_{reg}(z, z') H_{\mu\nu,\nu}(z')
1\!{\rm l}_{reg}(z', z') + O(\lambda^3) }\eqno(2.30)$$
Now we seem to have cancelled the operators of the form $\lapl X$, but the
operators  on the right hand side of eq. (2.30), which do not show any explicit
dependence on  $\lapl X$, still have to be normal ordered and expanded into a
set of local operators  as in our treatment of the term quadratic in $S_{int}$
above. In the case of the  operator $4\pi V_{\mu,\nu}(z)\tilde H_{\mu\nu}(z,
z)$, e.g., this expansion  looks like
$$\eqalign{&4\pi V_{\mu,\nu}(z)\tilde H_{\mu\nu}(z, z) =  k_0(\nabla,\nabla')
V_{\mu,\nu}(z) H_{\mu\nu}(z) + \cr &k_1(\nabla,\nabla') \lapl^{cov}_z \sigma(z)
V_{\mu,\nu}(z) H_{\mu\nu}(z) +\cr  &k_2(\nabla,\nabla') :V_{\lambda,\rho}(z)
H_{\lambda\rho,\mu\nu}(z)\partial^a X^\mu  \partial_a X^\nu:+ \cr
&k_3(\nabla,\nabla') :V_{\lambda,\nu}(z) H_{\lambda\nu,\mu} \lapl^{cov}X^\mu: +
...} \eqno(2.31)$$
where $\nabla$ and $\nabla'$ denote target space derivatives which act on the
first  and second factor, respectively, of the following product of operators.
(Of course,  the target space derivatives could also be written as p or p$'$
after Fourier  transformation.)  As above  the normal ordering :...: refers
only to contractions with at least one world sheet  derivative acting on the
propagator. The dots in eq. (2.31) stand for further  operators corresponding
to massive modes. The coefficients $k_i(\nabla,\nabla')$  are given by
integrals over operator
product coefficients similar to the $\tilde C_{i j}^k$, but with no Weyl
derivative involved. For instance,  $k_0(\nabla,\nabla')$ and
$k_1(\nabla,\nabla')$ are given by the contributions of order 0 and 1 in
${1\over R^2} $ (cf. below eq. (2.19)) to the integral
$\int d^2z'\sqrt{g(z')} 1\!{\rm l}_{reg}(z, z') 1\!{\rm l}_{reg}(z', z)
\exp {[-pp'G^{cov}_{reg}(z,z')]}$, where $-pp'$ becomes $\nabla \nabla '$ after
Fourier-transforming back to position space.
Because of the last term in eq. (2.31) it seems as if we have actually failed
to eliminate the operator $\lapl  X$. However, to order $\lambda^2$ one can
replace normal ordered operators $:A_\mu\lapl^{cov}X^\mu:$ by
$(1/2)\lapl^{cov}_z\sigma :A_{\mu,\mu}:$ for
any $A_\mu$ quadratic in the fields.
This can be seen by considering expectation values of such operators and using
 $$\lapl^{cov}_zG^{cov}(z,z')=-4\pi 1\!{\rm l}(z,z')-{1\over2}\lapl^{cov}_z
\sigma(z).\eqno(2.32)$$
as well as the target space momentum conservation enforced by the zero mode
integration.
(The cutoff should be put to zero in contractions with external sources,
comp. \lb 33\rb ).
The same kind of replacement should then of course be also
applied to the  second-order contributions $V^{(2)}_\mu$ to the
$\beta$-function of $\lapl X$ mentioned above. All these contributions have
carefully to be taken into account in the computation of the dilaton
$\beta$-function $\beta_\Phi$. Note, however, that after the expansion of the
right-hand side of eq. (2.30)  into local normal-ordered operators as in eq.
(2.31), we get contributions to the tachyon and graviton $\beta$ functions as
well. Since all these contributions to $\beta_\Phi, \beta_T$ and
$\beta_{H_{\mu\nu}}$ can be interpreted as the use of Schwinger-Dyson equations
on the world sheet, they will be called $\beta^{SD}$ subsequently. We remark
that they play a similar role as  the so-called diffeomorphism terms in the
loopwise expansion [5,36--41].

In the discussion of the linearized $\beta$ functions, which require only the
information about the propagator and its derivatives at coincident points,  we
will work in a general scheme and parametrize the unknown quantities.  On the
other hand, the coefficients $\tilde C_{m n}^{k (i)}$ of eq. (2.17)  which
become relevant at order $\lambda^2$ cannot be computed in an arbitrary
scheme; standard methods like the heat  kernel regularization do not lead to
tractable expressions. For our calculations we used
 $$G_{reg}(z_1,z_2)=-\ln(\vert z_1-z_2\vert^2+\epsilon^2e^{-{1\over2}\sigma
(z_1)}e^{-{1\over2}\sigma(z_2)}),\eqno(2.33a)$$
 $$G^{cov}_{reg}(z_1,z_2)=G_{reg}(z_1,z_2)-{1\over2}\sigma(z_1)-
{1\over2}\sigma(z_2) +\ln {\epsilon ^2}.\eqno(2.33b)$$
a form which has already been considered by Fradkin and Tseytlin [1].
In eqs.(2.33), we have made the cutoff dependence explicit to show that
also on the quantum level, it is only the combination $\epsilon^2 e^{-\sigma
 (z)}$ that enters into all expressions (cf. the remark below eq.(2.6)).
 Actually one has to check that the regularization (2.33a)
is reparametrization-invariant. Let us consider
the reparametrization Ward identity
 $$\eqalign{&
2\nabla^z\left({\delta\over{\delta g^{zz}}}G_{reg}(z_1,z_2)\right)=
\nabla_z\left({\delta\over{\delta \sigma(z)}}G_{reg}(z_1,z_2)\right)\cr
&+(1\!{\rm l}(z,z_1)\partial_{z_1}+1\!{\rm l}(z,z_2)\partial_{z_2})
G_{reg}(z_1,z_2)\cr}\eqno(2.34)$$
To evaluate ${{\delta G_{reg}(z_1,z_2)}\over{\delta g^{zz}}}$, we have to
specify an extension of (2.34) to a general gauge where $g_{zz}\not=0$. Thus
one has to show that this extension can be chosen such that (2.34) is
fulfilled.
Making the ansatz
 $$G_{reg}(z_1,z_2)=-\ln(\vert z_1-z_2\vert^2+\epsilon^2 e^{-\sigma/2}(z_1)
e^{-\sigma/2}(z_2)+h(z_1,z_2))\eqno(2.35)$$

with $h(z_1,z_2)\to 0$ for $g_{zz}\to 0$, we find the solution
 $$\eqalign{
h(z_1,z_2)=&\Bigl\{-{\epsilon^2\over8}e^{-1/2}(z_1)e^{-\sigma/2}(z_2)
[{\partial_1\over
\bar\partial_1}(\sqrt gg^{z_1z_1})+(1\leftrightarrow 2)]\cr
&-{1\over4}[(\bar z_1-\bar z_2){1\over\partial_1}(\sqrt g g^{z_1z_1})-(1
\leftrightarrow 2)]\Bigr\}+c.c.\cr}\eqno(2.36)$$
to first order in $g^{zz}$.

Clearly, Weyl variations of the propagator (2.33) at coincident points
with world sheet derivatives acting on it differ from expressions obtained
elsewhere
[42 -- 46]. Note in this context that even Weyl derivatives of
self-contractions
of the propagator involving only $\partial_z$ (or $\partial_{\bar z}$)
derivatives are not uniquely specified by reparametrization invariance, cf. the
appendix of ref. [46]. (These expressions are unique only under the assumption
that the limit $\epsilon \to 0$ of $
{\delta\over{\delta g^{zz}}}\partial^m_{z_1}\partial^n_{z_2}G_{reg}(z_1,z_2)$
is continuous at $z_1=z_2$, which is, however, never required in our
framework.)

At this point, we would like to make another comment on the literature. In
[12],
the following recipe for the extraction of the Weyl anomaly of the
Tachyon-Tachyon
operator product has been proposed (compare also [14])
 $$\eqalign{
(e^{-kk'G(z,z')})_{reg}=&-\sum^\infty_{n=0}{\pi\over{4^n(n!)^2}}{{\epsilon^
{2(kk'+n+1)}}\over{kk'+n+1}}\cr
&\cdot e^{-kk'G_{reg}(z,z')}[\lapl^n_z+C(z)]1\!{\rm l}(z,z')\cr}\eqno(2.37)$$
and similarly for products of operators containing world sheet derivatives.
Here $C(z)$ denotes local reparametrization-invariant terms, i.e. powers of the
two-dimensional curvature  and its derivatives. For the flat metric
$\sigma=const$, (2.37) reduces to
$$(e^{-kk'G(z,z')})_{reg}=-\sum^\infty_{n=0}{\pi\over{4^n(n!)^2}}{{\epsilon^{2
(kk'+n+1)}}\over{kk'+n+1}}\left({\lapl_z}\right)^n\delta(z-z')\eqno(2.38)$$
This is just the ``distributional Taylor expansion'' of
 $$e^{-kk'G(z-z')}\theta(\vert z-z'\vert-\epsilon)\equiv \vert z-z'
\vert^{2kk'}\theta(\vert z-z'\vert -\epsilon)\eqno(2.39)$$
into derivatives of the $\delta$-function. The analogous interpretation
of the covariant form (2.37) is obvious.
Note, however, that on a compact world sheet the two-point function at
noncoincident points is related to the one at coincident points by a Taylor
expansion (in the geodesic distance, which is bounded and therefore arbitrary
powers can be integrated). Therefore, the propagator and its derivatives at
coincident points cannot be specified independently of the choice (2.37), which
has, however, been done in [12, 14]. We do not expect to obtain consistent
$\beta$-functions
when mixing different regularizations in the same calculation.
Therefore we insist on using one and the same regularization for all possible
expressions involving the propagator and its derivatives at coincident or
noncoincident points. Then the ''$\theta$- function prescription'' (2.37)
(and its obvious generalizations involving derivatives) is in fact inadmissible
no matter how one specifies the selfcontractions, because it cannot be derived
from any regularized propagator.

\bigskip
3.)
Let us now turn to the resulting $\beta$ functions corresponding to the
operators
shown in (2.9). We start with considering the first order in $\lambda$,
i.e. taking only the terms
linear in the string modes $\phi = \lb T,H_{\mu \nu},\Phi \rb $ in eq.(2.4)
into account. The corresponding calculation requires the knowledge of the
regularized propagator and
some of its derivatives at coincident points, i.e. of expressions of the form
$G_{reg}(z,z') \bigr\vert_{z=z'}$, $\partial_{a} G_{reg}(z,z')
\bigr\vert_{z=z'}$, $\partial_{a} \partial_{a} G_{reg}(z,z')
\bigr\vert_{z=z'}$ and $\partial_{a} \partial_{a}' G_{reg}(z,z')
\bigr\vert_{z=z'}$. The first two expressions are fixed if one requires that
the regularization preserves general coordinate invariance on the world sheet
\lb 42-46 \rb:
$$\eqalign{G_{reg}(z,z') \bigr\vert_{z=z'} = \sigma(z),} \eqno(3.1a)$$
$$\eqalign{\partial_{a} G_{reg}(z,z') \bigr\vert_{z=z'} = {1\over 2}
\partial_{a} \sigma(z),} \eqno(3.1b)$$
The last two expressions involving mixed derivatives, on the other hand, depend
on the regularization scheme, which we can let arbitrary at this stage.
However,
they are related to each other if we assume
the Leibniz rule to hold. We then have
$$\eqalign{\partial_{a} \partial_{a} G_{reg}(z,z') \bigr\vert_{z=z'}
= {\beta \over 2}\partial_{a} \partial_{a}\sigma(z) +
\gamma e^{\sigma(z)},} \eqno(3.2a)$$
and
$$\eqalign{\partial_{a} \partial_{a}' G_{reg}(z,z') \bigr\vert_{z=z'}
= {{(1-\beta)} \over 2}\partial_{a} \partial_{a}'\sigma(z) +
\gamma e^{\sigma(z)},} \eqno(3.2b)$$
with $\beta, \gamma$ arbitrary.
\smallskip
With the ansatz (2.22) the resulting $\beta$ functions read
$$\beta^{(1)}_T = -T-{1\over2}T,_{\mu\mu} + \gamma H_{\mu \mu} + \gamma
\hat F_{\mu}^{(1)},_{\mu}  \eqno(3.3a)$$
$$\eqalign{&\beta^{(1)}_{H_{\mu\nu}} = -\Phi,_{\mu\nu}-{1\over2}H_{\mu\nu},_
{\rho\rho}-
{1\over2}H_{\rho\rho},_{\mu\nu}
+{1\over2}H_{\mu\rho},_{\rho\nu}+{1\over2}H_{\nu\rho},_{\rho\mu} +{{\beta}
\over 2}
H_{\rho\rho,\mu\nu}}
\eqno(3.3b)$$
$$\beta^{(1)}_{\lapl X^\mu} = -\Phi,_{\mu}-{1\over2}H_{\rho\rho},_{\mu}+
{1\over2}
H_{\mu\rho},_{\rho}+
{1\over2}H_{\rho\mu},_{\rho}-\hat F_{\mu}^{(1)}
 + {\beta \over 2} H_{\rho\rho, \mu} \eqno(3.3c)$$
$$\beta^{(1)}_{\Phi} = {{d-26}\over6}-{1\over2}\Phi,_{\rho\rho} +
{\beta \over 2}
\hat F_\mu^{(1)} , _\mu .\eqno(3.3d)$$
These expressions actually appear multiplied with one of the
operators of (2.9). Since the contributions to the $\beta$ functions of
$O(\lambda^2)$ will arise in the form of normal ordered operators below,
we will rewrite (3.3a) to (3.3d) in terms of normal ordered operators as well.
This does
not add any new information, but amounts to add linear combinations of eqs.
(3.3b) and (3.3c) to eqs. (3.3a) and (3.3d).
The normal ordering of operators of the type b) and c) of eq. (2.9) involves
contractions without Weyl derivatives, hence according
to the remark in section 2
we should perform these using the covariant propagator (2.33b).
Therefore we need
$$\eqalign{\partial_{a} \partial_{a} G_{reg}^{cov}(z,z') \bigr\vert_{z=z'}
= {{(\beta-1)} \over 2}\partial_{a} \partial_{a}\sigma(z) +
\gamma e^{\sigma(z)},} \eqno(3.4a)$$
and
$$\eqalign{\partial_{a} \partial_{a}' G_{reg}^{cov}(z,z') \bigr\vert_{z=z'}
= {{(1-\beta)} \over 2}\partial_{a} \partial_{a}\sigma(z) +
\gamma e^{\sigma(z)},} \eqno(3.4b)$$
 The final result becomes
particularly simple if expressed in terms of redefined tachyon - and
dilaton - fields
$$\tilde T = T - \gamma H_{\mu \mu} \eqno(3.5a)$$
$$\tilde \Phi = \Phi - {\beta \over 2} H_{\mu \mu} \eqno(3.5b)$$
and introducing
$$\Gamma_\rho = \tilde \Phi,_{\rho}+{1\over2}H_{\mu\mu},_{\rho}-{1\over2}
H_{\mu\rho},_{\mu}-{1\over2}H_{\rho\mu},_{\mu}. \eqno(3.5c)$$
Note that the redefinition (3.5a) can be understood as the result of the
change to normal ordered operators, since we have
$H_{\mu\nu}\partial_aX^\mu\partial^aX^\nu =$
$ :H_{\mu\nu}\partial_aX^\mu\partial^aX^\nu:
 - {{\beta-1}\over 2}\lapl\sigma H_{\mu\mu}-\gamma e^{\sigma(z)} H_{\mu\mu}$.
Eq. (3.5b) differs, however, by a factor 1/2 $H_{\mu\mu}$ from the naive
expectation.
For the $\beta$ functions we find
$$\beta_T^{(1)} = -\tilde T-{1\over2}\tilde T,_{\mu\mu}, \eqno(3.6a)$$
$$\beta^{(1)}_{H_{\mu\nu}} = \eqalign{&-\tilde \Phi,_{\mu\nu}
-{1\over2}H_{\mu\nu},_{\rho\rho}-{1\over2}H_{\rho\rho},_{\mu\nu}
+{1\over2}H_{\mu\rho},_{\rho\nu}+{1\over2}H_{\nu\rho},_{\rho\mu},}
\eqno(3.6b)$$
$$\beta^{(1)}_{\lapl X^\mu} = -\Gamma_\mu-\hat F_{\mu}^{(1)},\eqno(3.6c)$$
$$\beta^{(1)}_\Phi = {{d-26}\over6}-{1\over2}\tilde \Phi,_{\rho\rho}-{1\over 2}
(H_{\rho \rho},_{\mu \mu} - H_{\rho \mu},_{\mu \rho}) + {1 \over 2}(\hat
F_{\mu,\mu}^{(1)} + \Gamma_{\mu,\mu}) .\eqno(3.6d)$$
We see that the scheme dependence parametrized by $\beta$ and $\gamma$ is fully
removed by the field redefinitions (3.5). Furthermore, these redefinitions lead
to covariant expressions for $\beta_T$ and $\beta_{H_{\mu\nu}}$ to order
$\lambda^1$.
(Subsequently the twiddles on $\tilde T$ and $\tilde \Phi$ will be omitted for
simplicity.)
$\beta_\Phi^{(1)}$, on the other hand, is not covariant for a general $\hat
F^{(1)}_\mu$,  but certainly for the choice $\hat F^{(1)}_\mu = -\Gamma_\mu$
required to satisfy  eq. (3.6c). This is the first indication of the relevance
of the Schwinger - Dyson  terms (leading to the presence of  $\hat F^{(1)}_\mu$
in $\beta_\Phi$)  for target space covariance. We remark that a more cavalier
approach using  the noncovariant propagator $G_{reg}$ for normal ordering would
have produced the  same result except for the last term in $\beta_\Phi^{(1)}$.
However, working  with $G_{reg}$ instead of $G_{reg}^{cov}$ will not give
consistent results  when computing the contributions to $\beta_\Phi$ of order
$\lambda^2$ according  to the method described in section 2.
\smallskip
In the above analysis, we ignored the contributions of the massive modes
$M^n$ completely. This is trivially allowed at order $\lambda $, but one
may be worried about possible effects of the $M^n$ on the self-contraction Weyl
 anomaly at order  $\lambda ^2$ (it is clear that they cannot enter into
the quadratic term of eq. (2.4) to $O(\lambda ^2)$). However, it is easy to
convince oneself that the only effect of taking the massive modes into
account would be that redefinitions (3.5a), (3.5b) are augmented by additional
terms containing the massive fields, and thus we can consistently ignore them
after passing to the "twiddled" fields. \smallskip
Let us now discuss the
role of the redundant operator  $\lapl_z X^\mu$. Ignoring its redundancy
corresponds to ignoring the possibility  to use the world sheet equations
of motion or to redefine the  field $X^\mu$; it amounts  to putting $\hat
F^{(1)}_\mu =0$ identically.  Under these circumstances the equation
(3.6c)  becomes $\Gamma_\mu = 0$ which corresponds to fixing the gauge of
the  graviton $H_{\mu \nu}$ . (Except for the presence of the dilaton, or
for  $\Phi,_\mu = 0$, this gauge is the so-called harmonic gauge.)
Inserting  this gauge fixing condition into eq. (3.6b), the graviton
equation becomes  $-{1\over 2}H_{\mu \nu},_{\rho \rho} = 0$. Thus, if we
require local Weyl invariance  of the partition function without allowing
for redefinitions of the field  $X^\mu$, the resulting constraint on the
background fields is not  general coordinate covariant in the target
space, but the gauge is fixed  in a way which reduces the graviton to the
physical degrees of freedom  and allows to invert its linearized equation
of motion, i.e. to find  its propagator. (The importance of redundant
operators for the restoration  of internal symmetries in the context of
renormalisation in nonlinear $\sigma$  models has been observed by Wegner
\lb 47\rb \ before.) \smallskip
In the following we will, however, satisfy
eq. (3.6c) by  choosing $\hat F^{(1)}_\mu = -\Gamma_\mu$. In this way we
effectively pass  from the linearly dependent set of operators $e^{ikX}$,
$ e^{ikX}\partial X^\mu
 \partial X^\nu$, $e^{ikX}\lapl\sigma$, $e^{ikX}\lapl X^\mu$ to the
linearly  independent set $e^{ikX}$, $ e^{ikX}\partial X^\mu \partial
X^\nu$,
 $e^{ikX}\lapl\sigma$ and hence obtain conditions for local Weyl
invariance  which are not only sufficient but also necessary.
\smallskip
Let us now proceed to the computation of the OPC's $\tilde C^{k(i)}_{mn}$ of
eq. (2.17). For the $T, H_{\mu\nu},\Phi$ system considered below, only $i=1,2$
is relevant. The corresponding coefficients are obtained as indicated in
section 2, i.e. we write down the functional $\tilde C_{m,n}^k(z;p_1,p_2;g)$
for the constant curvature metric and expand to first order around the
flat-space limit $R=\infty$. More precisely, we have
 $$\tilde C^{k(1)}_{mn}(p_1,p_2)=\lim_{R\to\infty}\tilde C^k_{mn}
(z;p_1,p_2;g)\eqno(3.9)$$
and
$$\tilde C^{k(2)}_{mn}(p_1,p_2)={1\over2}{\partial\over
{\partial(\epsilon^2/R^2)}}\Bigr\vert_{R=\infty}\tilde C^k_{mn}
(z;p_1,p_2;g).\eqno(3.10)$$
where the derivative is understood to act under the integrals in
$\tilde C^k_{mn}$,
see eq. (2.16). In this way, all relevant OPC's are given in terms of
flat space integrals.

In addition to the contributions to the $\beta$-functions from the term
quadratic in $S_{int}$ in eq. (2.4), we have the contributions from the
right-hand side of eq. (2.30) after the expansion into local normal-ordered
operators. In the case
of the dilaton $\beta$ function $\beta_\Phi$, also the effects due to the
substitution of $\lapl X^\mu$ have to be taken into account. The contributions
from the term quadratic in $S_{int}$ will be denoted by $\beta^{OPE}$,
whereas the other kinds
of contributions will be called $\beta^{SD}$ subsequently. We present the
result
after Fourier-transforming back to position space (in the target space).
Accordingly the coefficients $a_i$ to $c_i$ below  are functions of
$x=\nabla\cdot \nabla'$ (as the coefficients $k_i$ of eq. (2.31)), where
$\nabla$ acts on the first factor, and $\nabla'$ on the second factor; their
$x$ dependence is given in the appendix. Altogether we find  for the $\beta$
functions in the normal-ordered basis:
 $$\eqalign{
\beta^{OPE}_T=&a_1H_{\mu\nu}H_{\mu\nu}+a_2H_{\mu\nu,\rho}H_{\mu\rho,\nu}
+a_3H_{\mu\rho,\nu\lambda}H_{\nu\lambda,\mu\rho}\cr
&+a_4 H_{\mu\nu}T_{,\mu\nu}+{1\over{16}}T^2\cr}\eqno(3.11)$$
 $$\eqalign{
\beta_T^{SD}=&-a_5\Phi T-a_6\Phi_{,\mu\nu}H_{\mu\nu}+a_4H_{\mu\rho,\rho}
T_{,\mu}\cr
&-{a_5\over 2}TH_{\rho\rho}+4a_3 H_{\mu\nu} H_{\mu\lambda,\lambda\nu}+a_3
H_{\rho\nu,\mu}H_{\mu\lambda,\lambda\nu\rho}\cr
&-{a_6\over2}H_{\mu\nu}H_{\lambda\lambda,\mu\nu}\cr}\eqno(3.12)$$
 $$\eqalign{
\beta^{OPE}_{H_{\mu\nu}}=&b_1H_{\lambda{\mu,\rho}} H_{\lambda\nu,\rho}-b_1
H_{\lambda\mu,\rho} H_{\rho\nu,\lambda}+b_1H_{\lambda \rho}
H_{\mu\nu,\lambda\rho}\cr
&+b_2H_{\lambda{\rho}} H_{\lambda\rho,\mu\nu}-b_3
H_{\lambda\rho,\nu} H_{\lambda\rho,\mu}\cr
&+b_4H_{\lambda\rho,\kappa}H_{\kappa\rho,\lambda\mu\nu}-b_5
H_{\sigma\rho,\nu\lambda}H_{\sigma\lambda,\rho\mu}\cr
&+b_6 H_{\lambda\rho,\kappa\sigma}H_{\kappa\sigma,\lambda\rho{\mu\nu}}
-2b_6 H_{\kappa\rho,\sigma\lambda\nu}H_{\sigma\lambda,
\kappa\rho\mu}\cr
&-b_7 H_{\lambda\rho}H_{\lambda(\mu,\nu)\rho}-b_8H_{\lambda
\rho(\nu}H_{\mu)\lambda,\rho}\cr
&+{b_7\over2}H_{\lambda\rho,\sigma(\mu}H_{\nu)\sigma,\lambda\rho}
-{b_7\over 2}H_{\lambda\rho,\sigma}H_{\sigma(\nu,\mu)\lambda\rho}\cr
&+{1\over8}TH_{\mu\nu}-b_9T_{,\lambda}  H_{\lambda(\mu,\nu)}\cr
&-b_{10}H_{\lambda\rho,(\mu}T_{,\nu)\lambda\rho}+b_{10}T_{,\lambda\rho}
H_{\lambda\rho,\mu\nu}\cr
&-b_{11}TT_{,\mu\nu}\cr}\eqno(3.13)$$
 $$\eqalign{
\beta^{SD}_{H_{\mu\nu}}=&-{b_1\over 4}\Phi T_{,\mu\nu}-b_{12}\Phi H_{\mu\nu}
+2b_1\Phi_{,\lambda}H_{\lambda(\mu,\nu)}\cr
&-b_{13}\Phi_{,\lambda\rho}H_{\lambda{\rho,\mu\nu}}+
b_{10}H_{\lambda{\rho,\rho}}
T_{,\lambda{\mu\nu}}\cr
&-{b_1\over 8}H_{\rho\rho}T_{,\mu\nu}-{{b_{12}}\over2}H_{\rho\rho}H_{\mu\nu}-
b_{14}H_{\lambda(\mu,\nu)}H_{\lambda\rho,\rho}\cr
&+b_1H_{\rho\rho,\lambda}H_{\lambda(\mu,\nu)} +b_1 H_{\lambda{\rho,\rho}}
H_{\mu\nu,\lambda}\cr
&+b_{15}H_{\lambda\rho,\mu\nu}H_{\lambda\sigma,\sigma\rho}-{{b_{13}}\over2}
H_{\lambda\rho,\mu\nu}H_{\sigma\sigma,\lambda\rho}\cr
&+2b_6H_{\lambda\sigma,\sigma\rho\kappa}H_{\kappa\rho,\lambda{\mu\nu}}-
b_7H_{\lambda
\sigma,\sigma\rho}H_{\rho(\mu,\nu)\lambda}\cr}\eqno(3.14)$$

 $$\eqalign{
\beta^{OPE}_\Phi=&c_1H_{\mu\nu}\Phi_{,\mu\nu}-
{c_1\over8}H_{\mu\nu}T_{,\mu\nu}\cr
&+{1\over8}T\cdot\Phi+{1\over{16}}TH_{\rho\rho}-c_2 T_{,\lambda}H_{\mu\lambda,
\mu}\cr
&+c_3T_{,\lambda\rho}H_{\lambda\rho,\mu\mu}+c_4H_{\mu\nu}H_{\mu\nu}+{c_1\over2}
H_{\mu\nu}H_{\rho\rho,\mu\nu}\cr
&+c_5 H_{\lambda\rho}H_{\lambda\rho,\mu\mu}-c_6H_{\mu\lambda,\nu}
H_{\mu\nu,\lambda}
\cr
&-c_7H_{\sigma\rho}H_{\sigma\mu,\rho\mu}-c_8H_{\kappa\lambda,\mu\nu}
H_{\mu\nu,\kappa\lambda}\cr
&-{c_7\over2}H_{\lambda\rho,\sigma}H_{\sigma\mu,\lambda\rho\mu}+
c_9H_{\rho\sigma,
\mu\mu\lambda}H_{\sigma\lambda,\rho}\cr
&+c_{10}H_{\kappa\rho,\sigma\mu\mu\lambda}H_{\sigma\lambda,\rho\kappa}
+c_{11}(T_{,\mu}T_{,\mu}+TT_{,\mu\mu})\cr}\eqno(3.15)$$

 $$\eqalign{
\beta^{SD}_\Phi=&{c_1\over8}\Gamma_\mu T_{,\mu}-c_1\Gamma_\mu\Gamma_\mu-
2c_9\Gamma_{\mu,\nu\rho}N_{\mu\nu\rho}\cr
&+c_{12}\Gamma_{\mu,\nu}H_{\mu\nu}-c_3(\Gamma_\lambda
T_{,\lambda\mu})_{,\mu}\cr
&+2c_7(\Gamma_{\lambda,\kappa}N_{\lambda\kappa\mu})_{,\mu}-4c_{10}
(\Gamma_{\lambda,\nu\rho}N_{\lambda\nu\rho,\mu})_{,\mu}\cr
&-c_{13}(\Gamma_{\lambda,\nu}H_{\lambda\nu,\mu})_{,\mu}\cr}\eqno(3.16)$$
where $\Gamma_\mu$ is given by eq. (3.5c) and $N_{\rho\mu\nu}$ by
$$N_{\rho\mu\nu}={1\over2} H_{\mu\nu,\rho}-H_{\rho(\mu,\nu)}.\eqno(3.17)$$
As usual brackets denote weighted symmetrization. The $T$ and $\Phi$ fields
are again the redefined ones on the left hand side of eqs. (3.5) where, for
the regularization (2.33a), $\beta = 1$ and $\gamma = -4$.

The coefficients $a_i$ to $c_i$ are of infinite order in $x$ ( or in $p\cdot
p'$ in momentum space), since our results are nonperturbative in the string
constant $\alpha'$ or the number of loops on the world sheet. Actually
these coefficients even have poles in $p\cdot p'$; as is easily checked,
however, these poles appear only in the unphysical region of the momenta.
They are the remnants after analytic continuation of the fact that the
integrals defining the $\tilde C^{k(i)}_{mn}$ diverge outside a certain range
for $pp'$. This divergence is actually expected because the
$\tilde C^{k(i)}_{mn}$ are independent of the 2d metric and therefore
contain no infrared regulator. On the other hand, the full operator
product - or the coefficient $\tilde C^k_{mn}(z,\sigma,p,p')$ before
discarding the nonlocal terms - should certainly be well-defined for any
$pp'$ since after all the theory is IR finite for any finite size of the
world sheet. Therefore what must - and does - happen is that the nonlocal
terms precisely cancel the infinities arising in the flat space integrals
for the $\tilde C^{k(i)}_{mn}$, resp. the poles which remain after analytic
continuation. While this cannot be seen in the simplest example eq.(2.19)
because its local part happens not to contain any pole in $pp'$,
it is a worthwhile
exercise to verify explicitly the cancellation for (2.19) with one of
the $T$'s replaced by $H_{\mu\nu}$. Even though our analytic continuation
prescription seems very natural from the point of view of string theory,
one should be careful therefore about attributing any deeper meaning to
the poles in $a_i, b_i, c_i$. The momentum dependence of these coefficients
far from the mass shells certainly depends strongly on the specific
structure of the regularization chosen. Hence one expects that it can be
changed by suitable field redefinitions which would connect our results with
those obtained in different schemes.

In order to compare our $\beta$ functions
with the ones obtained from string scattering
amplitudes, we have to impose the gauge conditions
$H_{\rho\rho}=H_{\mu\nu,\nu}=0$. Furthermore, the mass shell conditions $x=0$
for the $H \cdot \Phi$ and $H \cdot H$ terms resp. $x=1$ for the $T \cdot H$
terms and $x=2$ for the $T \cdot T$ terms have to be used in
$\beta_{H_{\mu\nu}}$, as well as $x=-1$ for the $H \cdot \Phi$ and $H \cdot H$
terms , $x=0$ for the $H \cdot T$ and $\Phi \cdot T$ terms, and $x=1$ for the
$T
\cdot T$ terms in $\beta_T$. In all cases one finds agreement and in this sense
our results coincide with the ones obtained by other methods [7 -- 14].
\bigskip
4.) Let us now discuss the structure of the $\beta$-functions. The operator
product expansion $V_a\cdot V_b\to V_c$ conserves powers of the $2d$ curvature,
i.e. if  either $V_a$ or $V_b$ contains $R^{(2)}(z)$, this will also hold for
$V_c$. Accordingly, the dilaton cannot enter the tachyon or graviton
$\beta$-functions via the OPE contributions, but only via the SD-terms.

As can be derived from superficial power counting, terms quadratic in
the tachyon $T$ would never appear in any of the $\beta$-functions to any
finite order in the number of loops on the world sheet; as is well known,
these are genuinely nonperturbative effects arising from the generation of
powers by the exponentiation of logarithms.

Note that, in particular in the case of the graviton $\beta$-function, we have
obtained a large number of terms which would vanish upon the use of gauge
conditions of the form $H_{\mu\nu,\mu}=H_{\mu\mu}=0$. These terms cannot
be seen if one obtains the $\beta$-functions via string scattering amplitudes.
On the other hand,
they are crucial for the discussion of manifest target-space
covariance.

The presence of terms involving $T$ or $\Phi$ without
derivatives acting on them
(as the $T\Phi$-term in $\beta_\Phi$) constitutes actually a paradox. In the
case of $\Phi$, the following simple argument shows that all $\beta$-functions
should depend only on derivatives of $\Phi$: Consider a  constant shift of
$\Phi,\Phi\to\Phi+a$, then the Gauss-Bonnet theorem on the sphere
 $${1\over{8\pi}}\int d^2z\sqrt gR^{(2)}=1\eqno(4.1)$$
leads to the following identity for the partition function $Z$:
 $${{\delta Z}\over{\delta a}}=Z.\eqno(4.2)$$
Considering the Weyl derivative of $Z$ one finds
 $${\delta\over{\delta a}}{{\delta Z}\over{\delta\sigma(z)}}=
{\delta\over{\delta a}}\langle\beta_iO_i(z)\rangle=\langle \beta_iO_i(z)\rangle
+\langle{{\delta \beta_i}\over{\delta a}}O_i(z)\rangle\eqno(4.3)$$
but also
 $${\delta\over{\delta\sigma(z)}}{{\delta Z}\over{\delta a}}=
{{\delta Z}\over{\delta
\sigma(z)}}=\langle \beta_iO_i(z)\rangle.\eqno(4.4)$$
Accordingly $\delta \beta_i/\delta a$ has to vanish for all $\beta_i$
(related to independent operators $O_i$).A similar argument in the
case of constant shifts of the tachyon $T$ (on a world sheet with finite
volume) shows that the $\beta$-functions should depend on the
tachyon zero mode only through a term $-T$ in $\beta_T$ (see also [48]).

In order to resolve the apparent contradiction with our results (and the ones
obtained by other authors), we have to remember that we used the Fourier
expansion eq. (2.5) to evaluate the nonperturbative contributions to the
$\beta$
functions. Thus the results can be trusted only for field configurations
$\varphi_i=T,h_{\mu\nu},\Phi$ which are Fourier-integrable. But if $\varphi_i
(x)$ falls off rapidly enough at infinity to admit a Fourier expansion,
$\varphi_i+a$ does not. Hence the constant shift which leads to the apparent
inconsistency cannot actually be performed. What the argument (4.1)-(4.4) (and
its analog for $T$) really tells us is how to extend our results from
configurations $\hat T,
\hat \Phi$, which are Fourier-integrable, to the shifted configurations
$T=\hat T+a\quad \Phi=\hat\Phi+b$ ($a,b$ constant) [29].
For example, the term ${1\over8}Th_{\mu\nu}$ in $\beta_{H_{\mu\nu}}$
should be replaced by ${1\over8}\hat T h_{\mu\nu}={1\over8}(T-a)h_{\mu\nu}$
with $a$ arbitrary. It is not clear at present, however, how to extend
the analysis to arbitrary configurations of $T$ and $\Phi$
(e.g. polynomials in $X$).
(For the special case of a linear dilaton $\Phi=Q^\mu X_\mu$, however,
a suggestion was made in [33].)

Let us now turn to the discussion of target space covariance and the
possibility
of relating the $\beta$ functions to equations of motion derived from a
covariant effective action. In contrast to the loopwise computation of
$\beta$-functions (associated with renormalizable interactions or massless
string backgrounds) employing the manifestly covariant background field method
[1 -- 4], the weak field expansion induces a breaking of target space
covariance through the  splitting $G_{\mu\nu}=\delta_{\mu\nu} +H_{\mu\nu}$. The
$\delta_{\mu\nu}$ part corresponds to the free kinetic term of the action which
is modified in the process of the regularization of the propagator, whereas the
$H_{\mu\nu}$ part is left unmodified and treated as  interaction. Target space
covariance should be restored, however, after suitable scheme-dependent field
redefinitions
 $$\tilde\varphi_i=\varphi_i+f_i(\varphi).\eqno(4.5)$$
Actually, covariance also seems to be broken by the implicit
use of the flat path integral measure $DX$ instead of the covariant one
 $$DX_{cov}=DX\sqrt{det\ G_{\mu\nu}}.\eqno(4.6)$$
On the other hand, at least in the weak field expansion the factor $\sqrt{det\
G_{\mu\nu}}$ can be expanded in $H_{\mu\nu}$, exponentiated and added to the
interaction $S_{int}$ [57]. Since we started with the most general local
action, this should thus amount to just another field redefinition.
Nevertheless, it seems advisable to take an open-minded view on covariance
and the way it may be realized in the presence of nonperturbative effects.
Our results below suggest that the $\beta$ functions
are not simply given by the variation $\delta S_{eff}(\varphi)/\delta\varphi_i$
of an underlying covariant target space action $S_{eff}(\varphi)$, but satisfy
a relation of the type
 $${{\delta S_{eff}^{(\varphi)}}\over{\delta\varphi_i}}=\kappa^{ij}\beta_j.
\eqno(4.7)$$
$\kappa^{ij}$ should play the role of a metric in the space of operators of the
theory. This expectation is a consequence of the so-called $c$-theorem [49 --
52],  which is, however, generally formulated in the finite-dimensional space
of renormalizable interactions.

A comprehensive analysis of target-space convariance and relations of the type
(4.7) to
$O(\lambda^2)$ and infinite order in space-time derivatives would go beyond
the scope of the present paper. To first order in $\lambda$, however, we
arrived
(after the simple field redefinitions (3.5)) at the covariant expressions (cf.
eqs. (3.6))
 $$\beta_T^{(1)}=-{1\over2}(\lapl+2)T\eqno(4.8a)$$
 $$\beta_{H_{\mu\nu}}^{(1)}=-\nabla_\mu\nabla_\nu\Phi-R^{(1)}_{\mu\nu}
\eqno(4.8b)$$
 $$\beta^{(1)}_\Phi={{d-26}\over6}-{1\over2}\lapl \Phi-{1\over2}R^{(1)}
\eqno(4.8c)$$
with
 $$R^{(1)}_{\mu\nu}={1\over2}H_{\mu\nu,\lambda\lambda}+{1\over2}
H_{\lambda\lambda
,\mu\nu}-{1\over2}H_{\lambda\mu,\nu\mu}-{1\over2}H_{\lambda\nu,\mu\lambda}.
\eqno(4.9)$$
These results are well known [1--4, 6] and can be related to the equations of
motion derived from a covariant target space
effective action (to the appropriate order in the weak field expansion), which
reads with our normalizations
 $$S_{eff}(\varphi)=\int d^Dx\sqrt G e^{-\Phi}[-R-(\nabla\Phi)^2+{1\over{64}}
(\nabla T)^2-{1\over{32}}T^2+{1\over{768}}T^3]\eqno(4.10)$$

To second order in the weak field expansion, the high orders in target space
derivatives within the coefficients $a_i, b_i, c_i$ of the $\beta$ functions
[cf. the appendix]
are evidently far from exhibiting manifest covariance. On the other hand, the
amount of freedom in the choice of field redefinitions (4.5) and the $\kappa$
function of (4.7) is considerable. Therefore we will only discuss some
selected items, typically by considering low orders in the target space
derivatives.

Let us first turn to the tachyon $\beta$ function $\beta_T$. The terms
involving just $T$ or $T\cdot H$ become to second order in the derivatives
$$\beta_T = -T-{1\over2}T_{,\mu\mu}+{1\over2}H_{\mu\nu}T_{,\mu\nu}+
{1\over2}H_{
\mu\rho,\rho}T_{,\mu}-{1\over4}H_{\rho\rho,\mu}T_{,\mu}+{1\over{16}}T^2+...
\eqno
(4.11)$$

We see that the $T\cdot H$ terms combine nicely to generate the covariant
Laplacian $\lapl^{cov}$ acting on $T$. Note that the SD-contribution (3.12) is
crucial in this respect.  The above terms agree with $16\cdot \delta
S_{eff}(\varphi)/\delta T$ with $S_{eff}$ given by (4.10).

Next we turn to the $H\cdot\Phi$ and $H\cdot H$ terms in $\beta_T$. We remark
that an expansion in target space derivatives is actually not well defined in
the case of $\beta_T$, because any term $\Delta$ quadratic in the fields can be
turned into $-{1\over2}\Delta_{,\mu\mu}$ by  a tachyon redefinition $T\to
T+\Delta$. In particular this is the way how  to deal with the term
$H_{\mu\nu}H_{\mu\nu}$ without derivatives in (3.11), which cannot be related
to a covariant expression otherwise. Tseytlin [29] has argued that the $H\cdot
H$ terms in $\beta_T$ motivate the appearance of a $TR_{\mu\nu\rho\sigma}R^{\mu
\nu\rho\sigma}$ term in $S_{eff}(\varphi)$. His suggestion  was based on
a direct comparison with string scattering amplitudes, which are obtained
under the assumption $H_{\mu\nu,\mu}=H_{\mu\mu} =0$ and on the corresponding
mass shells. Our framework allows to study these terms far from the mass shells
and to ask whether they can still be related to a covariant effective action.
To this end we expand up to fourth order in
the derivatives and perform the following tachyon redefinition:
 $$T\to T+{2\over3} H_{\mu\nu}H_{\mu\nu}-{{13}\over{18}}
H_{\mu\nu,\rho}H_{\mu\nu,\rho}+{1\over3}H_{\mu\nu,\rho}
H_{\mu\rho,\nu}.\eqno(4.12)$$
Then one obtains, after some algebra, the following $H\cdot\Phi$ and $H\cdot H$
terms:
 $$\eqalign{
\beta_T=...&-{4\over3}\Phi_{,\mu\nu}H_{\mu\nu}+{7\over9}\Phi_{,\mu\nu\rho}H_{\mu
\nu,\rho}-{4\over3}R^{(1)}_{\mu\nu}H_{\mu\nu}\cr +&{{13}\over9}
R^{(1)}_{\mu\nu,\rho}H_{\mu\nu,\rho}-{2\over3} R^{(1)}_{\mu\rho,
\nu}H_{\mu\nu,\rho}-{2\over3}R^{(1)}_{\mu\nu}H_{\mu\nu,\rho\rho}+...\cr}\eqno
(4.13)$$
One finds that they can be combined into
 $$\eqalign{
\beta_T=&...+{4\over3}H_{\mu\nu}\beta_{H_{\mu\nu}}^{(1)}+H_{\mu\nu,\rho}(
{2\over3}(\beta_{H_{\mu\rho}}^{(1)})_{,\nu}-{{13}\over9}
(\beta_{H_{\mu\nu}}^{(1)})_{,\rho})
\cr
&-{2\over3}R^{(1)}_{\mu\nu}H_{\mu\nu,\rho\rho}+...\cr}\eqno(4.14)$$
Since the first three terms involve the complete lowest order graviton $\beta$
function, which is equivalent to $\delta S_{eff}(\varphi)/\delta H_{\mu\nu}$
to this order, they are an obvious sign of a non-trivial $\kappa$ function
of eq. (4.7). Note, however, that the $\kappa$ function has to be a functional
containing target space derivatives, as also found in [53--55]. In the last
term in (4.14) the
$\Phi$ dependence allowing to promote $R_{\mu\nu}^{(1)}$ to the lowest
order graviton $\beta$ function is missing, furthermore it does not exhibit
an index structure allowing to relate it to the proposal of Tseytlin.
Remember again that the term could be equipped with additional
derivatives by an appropriate additional tachyon redefinition, thus it does
not lead to a contradiction to the conjecture (4.7). The possible form of an
effective action which reproduces the string scattering amplitudes and
simultaneously allows to obtain the $\beta$ functions via (4.7) far off shell
thus remains an open problem.

In the case of the dilaton and graviton $\beta$  functions we restrict
ourselves
to the $H\cdot H,H\cdot\Phi$ and $\Phi\cdot\Phi$ terms, where an expansion
in target space derivatives does make sense.
To second order in derivatives, the corresponding terms in the dilaton $\beta$
function are
$$\eqalign{
\beta_\Phi=&-\Phi_{,\mu\mu}-{1\over2}R^{(1)}-{1\over2}R^{(2)}-{1\over6}
H_{\mu\nu}R_{\mu\nu}^{(1)}-{1\over6}(H_{\mu\nu}H_{\mu\nu})_{,\lambda\lambda}\cr
&+{5\over6}H_{\mu\nu}\Phi_{,\mu\nu}+\Phi_{,\mu}(H_{\mu\lambda,\lambda}-{1\over2}
H_{\lambda\lambda,\mu})-{1\over2}\Phi_{,\mu}\Phi_{,\mu}+...\cr}\eqno(4.15)$$
After a dilaton redefinition
 $$\Phi\to\Phi-{1\over6}(H_{\mu\nu}H_{\mu\nu})\eqno(4.16)$$
they can be written as
 $$\beta_\Phi=-\lapl^{cov}\Phi-{1\over2}R-{1\over2}(\nabla\Phi)^2+{1\over6}
H_{\mu\nu}\beta^{(1)}_{H_{\mu\nu}}+...\eqno(4.17)$$
The first three terms, to second order in the fields, are the manifestly
covariant ones expected from the action (4.10), whereas the last one
is again a sign of a non-trivial $\kappa$ function (4.7).

In the graviton $\beta$ function the terms involving just $H$ and $\Phi$
are, to second order in derivatives,
 $$\eqalign{
\beta_{H_{\mu\nu}}=&-R^{(1)}_{\mu\nu}-R^{(2)}_{\mu\nu}-\Phi_{,\mu\nu}
-{1\over2}\Phi_{,\lambda}H_{\mu\nu,\lambda}+\Phi_{,\lambda}
H_{\lambda(\mu,\nu)}\cr
&-{1\over6}(H_{\lambda\rho}H_{\lambda\rho})_{,\mu\nu}+...\cr}\eqno(4.18)$$
After performing the dilaton redefinition (4.16), we find the expected result
$$\beta_{H_{\mu\nu}}=-R-\nabla_\mu\nabla_\nu\Phi+...\eqno(4.19)$$

Herewith we conclude the preliminary discussion of the relation between the
$\beta$ functions and a target space effective action. Evidently our results
(3.11) to (3.16) contain a large amount of additional information, in
particular
concerning the higher derivatives in target space. On the other hand, it is
known already from the graviton/dilaton system to fourth order in the
derivatives that it is quite complicated to disentangle the scheme dependence
or the freedom to perform field redefinitions from true physical effects in
general [53, 55].

We close this chapter with some remarks on the Curci-Paffuti relation. In
the context of renormalizable $\sigma$ models it is well known that the
dilaton $\beta$ function $\beta_\Phi$ has to be constant at a fixed point where
the other $\beta$ functions vanish [5, 6]. This condition is necessary in order
to allow for an interpretation of $\beta_\Phi$, at a fixed point, as
central charge. More explicitly, the following relation has been shown
(we present its leading order form only):
 $$\nabla_\nu\beta
_\Phi=\nabla^\mu\beta_{H_{\mu\nu}}+\nabla^\mu\Phi\beta_{H_{\mu\nu}}\eqno(4.20)$$
An elegant two-line argument of Polchinski (cited by De Alwis [52]) shows that
the constancy of the dilaton $\beta$ function follows from the vanishing of the
others even
nonperturbatively. The argument does not allow, however, to extract the
explicit
form of the generalization of the relation (4.20). In order to establish a
relation of this type for the $\beta$ functions including the tachyon, we will
proceed as follows: First we order the explicit expression for $\nabla_\nu
\beta_\Phi,\nabla^\mu\beta_{H_{\mu\nu}}$ and $\nabla^\mu\Phi\beta_{H_{\mu\nu}}$
according to their field content and start by considering the terms involving
the graviton and the dilaton only. Whereas eq. (4.20) obviously holds for the
part linear in the fields, there are so many quadratic terms that we have to
restrict ourselves to an expansion in the number of space-time derivatives
again. Then, to second order in the derivatives, we can use the results  eqs.
(4.15) and (4.18) to find
 $$\nabla_\nu\beta_\Phi=\nabla^\mu\beta_{H_{\mu\nu}}+\nabla^\mu\Phi\beta
_{H_{\mu\nu}}+{1\over6}(H_{\mu\rho}\beta_{H_{\mu\rho}})_{,\nu}\eqno(4.21)$$
for all $H\cdot H, H\cdot\Phi$ and $\Phi\cdot\Phi$ terms simultaneously.
Now let us turn to the terms involving the tachyon. Here we were able to
find a relation valid to all orders in the space-time derivatives. It requires
to add terms of the form $T\cdot\beta_T,T\cdot\beta_H$ and $H\cdot\beta_T$
to right-hand side of eq. (4.21), which leaves their validity in the pure
graviton/dilaton sector (to second order in the fields) untouched. The
relation which now includes all $T\cdot T,T\cdot H$ and $T\cdot\Phi$ terms
reads explicitly
 $$\eqalign{
\nabla_\nu\beta_\Phi=&\nabla^\mu\beta_{H_{\mu\nu}}+\nabla^\mu\Phi\beta_{H_{\mu
\nu}}+{1\over6} (H_{\mu\rho}\beta_{H_{\mu\rho}})_{,\nu}\cr
&-b_{11}(T\beta_{T,\nu}-\beta_TT_{,\nu})-b_2T_{,\mu}\beta_{H_{\mu\nu}}\cr
&+b_{10}(T_{,\lambda\mu}(\beta_{H_{\lambda\mu}})_{,\nu}-T_{,\lambda\mu\nu}
\beta_{H_{\lambda\mu}})\cr
&+{b_1\over8}(H_{\rho\rho,\nu}\beta_T-H_{\rho\rho}\beta_{T,\nu})+b_{10}(
H_{\lambda{\rho\rho}}\beta_{T,\lambda\nu}-\cr
&-H_{\rho\lambda,\rho\nu}\beta_{T,\lambda})-b_{10}H_{\rho\lambda,\nu}\beta_
{T,\lambda\rho}\cr
&+{b_1\over4}(\Phi_{,\nu}\beta_T-\Phi\beta_{T,\nu}).\cr}\eqno(4.22)$$
The existence of such a relation, although valid only to second order in
the fields (and to second order in the derivatives in the $H-\Phi$ sector),
provides a rather nontrivial check
of our results eqs. (3.11) to (3.16), in particular of the
tachyon dependence of $\beta_\Phi$ and $\beta_H$, which we have not discussed
before. The coefficients $b_i$ are the same ones as in the appendix. The fact
that these coefficients appear in (4.22)
and seemingly cannot be changed by
any reasonable field redefinitions
comes somewhat as a surprise to us.
Since we expect these coefficients to
contain noncovariant artefacts of our specific regularization scheme,
it follows that also
the explicit form of the analog of the Curci-Paffuti relation, generalized
to the inclusion of nonperturbative effects and massive modes as the tachyon,
does not appear to have a scheme-independent meaning.
Although not in contradiction to the argument given by
Polchinski, we thus have found only a weak form of the Curci-Paffuti theorem
- the constancy of $\beta_\Phi$ at the fixed point - but not a general relation
 between the $\beta $ functions
transforming covariantly under redefinitions of the couplings. However,
the existence of the relation (4.22) suggests that our $\beta$ functions,
to all orders in spacetime momenta, can be derived from an underlying
action with some symmetry which may be a deformed version of ordinary
general covariance, reducing to it only in the perturbative limit.

\bigskip
5.) Let us now summarize our results and add some conclusions. We have shown
that a nonperturbative analysis of local Weyl invariance on a curved world
sheet, including massive modes such as the tachyon, can be performed
explicitly.
It requires the treatment of some technical subtleties. First, the proper
computation of operator products into curvature-dependent operators was seen
to necessitate  the use of manifestly covariant Green functions on the compact
world sheet. Furthermore, we found that the elimination of redundant operators
also involves  nonperturbative effects, which are again represented in the
form of covariant operator products.
In order to obtain expressions for the $\beta$ functions which are local on the
world sheet, we proposed to discard contributions to these operator
products which are nonanalytic in the cutoff, with the motivation that
such terms should represent infrared rather than ultraviolet
effects. This is certainly an additional input to our framework which
deserves a deeper understanding. In particular, the implications of this
prescription for covariance should be investigated; for example,
the reason why our
beta functions  -contrary to naive expectation-
seem to be covariant only modulo terms which vanish
at the conformal point (cf. eqs.(4.14) and (4.17)) could be buried here.

In our analysis of the covariance properties of the $\beta $ functions and
their
relation to a covariant effective action it was essential that we did not
restrict ourselves to any gauge choice for the graviton or the dilaton.
We found that after suitable field redefinitions and a suitable form
of the $\kappa$ function, the $\beta$ functions are related to the simple
effective action (4.10). This has only been checked, however, to low orders in
target space derivatives and not for all possible combinations of fields. Here
further investigations are clearly desirable: The possible field redefinitions
and the amount of freedom in the $\kappa$ function should be studied
systematically in order to see whether further terms in the space-time
effective action are required, in particular concerning the tachyon/dilaton
sector and higher orders in the derivatives. Our results have already shown,
however, that the $\kappa$ function will involve space-time derivatives itself.
It is an open question at present whether there exists a systematic way to
derive the explicit  form of the $\kappa$ function.

We have established an explicit relation of the Curci-Paffuti type including,
however, the tachyon $\beta$ function. Such a relation is necessary in order to
be able to interpret the dilaton $\beta$ function as central charge as usual.
It would be desirable, though, to find the underlying principle from which such
relations can be derived nonperturbatively.

Let us add some remarks on further possible work along the lines of this paper.
It is clear to us that a treatment of the next order in the weak field
expansion would require an extension of our techniques, perhaps along the lines
of [56]. The essential difference to the second-order analysis is that the
backreaction of the massive background on the $\beta$ functions of the massless
or tachyonic modes can no longer be ignored, and hence the naive extension of
the above analysis would lead to an infinite system of coupled equations.
On the other hand, a treatment of a finite number of arbitrary massive
backgrounds, at the same level as our treatment of the tachyon to
$O(\lambda^2)$, should be possible using the techniques described in this
paper. We remark that in general there will appear more redundant
operators in addition to $\lapl X$ (e.g. $(\lapl X)^2$) which all have
to be eliminated simultaneously. Although certainly extensive, such a work
would shed some light on the realization of the stringy gauge symmetries
associated with these massive modes.

\bigskip \bigskip
{\bf Acknowledgement}\hfill \break
It is a pleasure to thank Prof. Tseytlin for helpful discussions and comments.
Also we are grateful to Christian Schubert for performing extensive
checks of our calculations.
\vfill\eject

{\bf Appendix}\hfill \break
\bigskip
We shall specify here the coefficients $a_i$ to $c_i$ appearing in the
second-order contributions to the $\beta$ functions eqs. (3.11) to (3.16).
As described in the text, they are functions of $x=\nabla\cdot\nabla'$, where
$\nabla$ acts on the first and $\nabla'$ on the second factor.
\bigskip

$$\matrix{
 a_1={{4+3x+x^2}\over{(2+x)(3+x)}}\hfill
 &a_2={{2(1-x)}\over{(2+x)(3+x)}}\hfill
 &a_3={2\over{(2+x)(3+x)}}\hfill
\cr
& & \cr
 a_4={1\over{2(1+x)}}\hfill
 &a_5={x\over{2(1+x)}}\hfill
 &a_6={{2(4+x)}\over{(3+x)(2+x)}}\hfill
\cr
& & \cr  & & \cr
 b_1={1\over{2(1+x)}}\hfill
 &b_2={{4+3x+x^2}\over{4(1+x)(2+x)(3+x)}}\hfill
 &b_3={{(1-x)}\over{2(1+x)(2+x)(3+x)}} \hfill
\cr
& & \cr
 b_4={{(1-x)}\over{2(1+x)(2+x)(3+x)}}\hfill
 &b_5={{(5-x)}\over{2(1+x)(2+x)(3+x)}} \hfill
 &b_6={1\over{2(1+x)(2+x)(3+x)}}\hfill
\cr
& & \cr
 b_7={2\over{(1+x)(2+x)}} \hfill
 &b_8={x\over{(1+x)(2+x)}}\hfill
 &b_9={1\over{4x}}\hfill
\cr
& & \cr
 b_{10}={1\over{8x(1+x)}}\hfill
 &b_{11}={1\over{64(1-x)}}\hfill
 &b_{12}={x\over{2(1+x)}}\hfill
\cr
& & \cr
 b_{13}={{7+2x}\over{(1+x)(2+x)(3+x)}}\hfill
 &b_{14}={{2-x}\over{(1+x)(2+x)}}\hfill
 &b_{15}={{7+x}\over{(1+x)(2+x)(3+x)}}\hfill
 \cr
  & & \cr   & & \cr
 c_1={1\over{2(1+x)}}\hfill
 &c_2={1\over{8x}}\hfill
 &c_3={1\over{16x(1+x)}}\hfill
\cr
& & \cr
 c_4={{x(2+5x+x^2)}\over{8(1+x)(2+x)(3+x)}}\hfill
 &c_5={{4+3x+x^2}\over{8(1+x)(2+x)(3+x)}}\hfill
 &c_6={{6+9x+x^2}\over{4(1+x)(2+x)(3+x)}}\hfill
\cr
& & \cr
 c_7={1\over{(1+x)(2+x)}}\hfill
 &c_8={x\over{4(1+x)(2+x)(3+x)}}\hfill
 &c_9={{1-x}\over{(1+x)(2+x)(3+x)}}\hfill
\cr
& & \cr
 c_{10}={1\over{4(1+x)(2+x)(3+x)}}\hfill
 &c_{11}={1\over{128(1-x)}}\hfill
 &c_{12}={1\over{3+x}}\hfill
\cr
& & \cr
 c_{13}={1\over{2(2+x)(3+x)}} \hfill & &
 \cr   }$$

\vfill\eject
\centerline{\bf References}

\item{1.} E. Fradkin and A. A. Tseytlin, Phys. Lett. {\bf 158B} (1985) 316;
Nucl. Phys. {\bf B261} (1985) 1
\item{2.} C. Callan, D. Friedan, E. Martinec, and M. Perry, Nucl. Phys.
{\bf B262} (1985) 593
\item{3.} A. Sen, Phys. Rev. {\bf D32} (1985) 2102; Phys. Rev. Lett.
{\bf 55} (1985)
1846;\hfill\break
C. Lovelace, Nucl. Phys. {\bf B273} (1986) 413
\item{4.} A. A. Tseytlin, Int. J. Mod. Phys. {\bf A4} (1989) 1257, and refs.
therein
\item{5.} G. Curci and G. Paffuti, Nucl. Phys. {\bf B286} (1987) 399;
\hfill\break
A. A. Tseytlin, Nucl. Phys. {\bf B294} (1987) 383
\item{6.} C. Callan, I. Klebanov, and M. Perry, Nucl. Phys. {\bf B278}
(1986) 78
\item{7.} S. Das and B. Sathiapalan, Phys. Rev. Lett. {\bf 56} (1986) 2664,
and ibid. {\bf 57} (1986) 1511
\item{8.} R. Akhoury and Y. Okada, Phys. Lett. {\bf 198B} (1987) 65; Nucl.
Phys. {\bf B318} (1989) 176
\item{9.} S. Jain, G. Mandal, and S. Wadia, Phys. Rev. {\bf D35} (1987)
3116
\item{10.} C. Itoi and Y. Watabiki, Phys. Lett. {\bf 198B} (1987) 486
\item{11.} R. Brustein, D. Nemeschansky, and S. Yankielowicz, Nucl. Phys.
{\bf B301} (1988) 224
\item{12.} Y. Watabiki, Z. Phys. {\bf C38} (1988) 411
\item{13.} I. Klebanov and L. Susskind, Phys. Lett. {\bf 200B} (1988) 446
\item{14.} Y. Watabiki, Phys. Lett. {\bf 210B} (1988) 113
\item{15.} J. Labastida and M. Vozmediano, Nucl. Phys. {\bf B312} (1989) 308
\item{16.} K. Sakai, Nucl. Phys. {\bf B326} (1989) 237
\item{17.} B. Sathiapalan, Nucl. Phys. {\bf B326} (1989) 376
\item{18.} T. Ohya and Y. Watabiki, Mod. Phys. Lett. {\bf A4} (1989) 543
\item{19.} S. Jain and A. Jevicki, Phys. Lett. {\bf 220B} (1989) 379
\item{20.} N. Hayashi, Phys. Lett. {\bf 238B} (1990) 50
\item{21.} T. Banks and E. Martinec, Nucl. Phys.
{\bf B294} (1987) 733
\item{22.} S. Das, Phys. Rev. {\bf D38} (1988) 3105
\item{23.} A. Redlich, Phys. Lett. {\bf 213B} (1988) 285
\item{24.} U. Ellwanger and J. Fuchs, Nucl. Phys. {\bf B312} (1989) 95
\item{25.} J. Hughes, J. Liu, and J. Polchinski, Nucl. Phys. {\bf B316}
 (1989) 15
\item{26.} A. A. Tseytlin, Int. J. Mod. Phys. {\bf A4} (1989) 4249
\item{27.} K. Sakai, Osaka preprint OU-HET 127(1989)
\item{28.} A. A. Tseytlin, Phys. Lett. {\bf 185B} (1987) 59
\item{29.} A. A. Tseytlin, Phys. Lett. {\bf 264B} (1991) 311
\item{30.} A. Cooper, L. Susskind, and L. Thorlacius, Nucl. Phys. {\bf B363}
 (1991) 132
\item{31.} G. Mandal, A. Sengupta, and S. Wadia, Mod. Phys. Lett.
{\bf A6} (1991) 543
1685
\item{32.} S. De Alwis and J. Lykken, Phys. Lett. {\bf 269B} (1991) 264
\item{33.} U. Ellwanger and J. Schnittger, Int. J. Mod. Phys. {\bf A7}
(1992) 3389
\item{34.} S. Arakelov, Math. USSR Izv. {\bf 8} (1974) 1167;\hfill\break
E. D'Hoker and D. Phong, Commun. Math. Phys. {\bf 124} (1989) 629
\item{35.}
E. D'Hoker and D. Phong, Rev. Mod. Phys. {\bf 60} (1988) 917
\item{36.} D. Friedan, Ann. Phys. {\bf 163} (1985) 318
\item{37.} S. De Alwis, Phys. Rev. {\bf D34} (1986) 3760
\item{38.} B. Fridling and A. van de Ven, Nucl. Phys. {\bf B268}
 (1986) 719
\item{39.} C. Hull and P. Townsend, Nucl. Phys. {\bf B274} (1986) 349
\item{40.} A. A. Tseytlin, Phys. Lett. {\bf 178B} (1986) 34
\item{41.} G. Shore, Nucl. Phys. {\bf B286} (1987) 349
\item{42.} S. De Alwis, Phys. Lett. {\bf 168B} (1986) 59
\item{43.} I. Ichinose and B. Sakita, Phys. Lett. {\bf 175B} (1986) 423
\item{44.} E. D'Hoker and D. Phong, Phys. Rev. {\bf D35} (1987) 3890
\item{45.} J. Polchinski, Nucl. Phys. {\bf B289} (1987) 465
\item{46.} Y. Tanii and Y. Watabiki, Nucl. Phys. {\bf B316} (1989) 171
\item{47.} F. Wegner, Z. Phys. {\bf B78} (1990) 33
\item{48.} T. Banks, Nucl. Phys. {\bf B361} (1991) 166
\item{49.} A. B. Zamolodchikov, JETP Lett. {\bf 43} (1986) 731;
           Sov. J. Nucl. Phys. {\bf 46} (1987) 109;\hfill\break
           A. Polyakov, Physica Scripta {\bf T15} (1987) 191
\item{50.} A. A. Tseytlin, Phys. Lett. {\bf 194B} (1987) 63;\hfill\break
           H. Osborn, Nucl. Phys. {\bf B308} (1988) 629;\hfill\break
           N. Mavromatos and J. Miramontes, Phys. Lett. {\bf 212B} (1988) 33
\item{51.} A. Cappelli, D. Friedan and J. Latorre, Nucl. Phys.
{\bf B352} (1991)
           616
\item{52.} S. De Alwis, Phys. Lett. {\bf 217B} (1989) 3467
\item{53.} I. Jack, D. R. T. Jones, and D. A. Ross,
 Nucl. Phys. {\bf B307} (1988) 130
\item{54.} U. Ellwanger, J. Fuchs, M. G. Schmidt,
 Nucl. Phys. {\bf B314} (1989) 175
\item{55.} M. C. Bento, O. Bertolami, and J. C. Romao,
Phys. Lett. {\bf 252B} (1990) 401
\item{56.} U. Ellwanger, Phys. Lett. {\bf 243B} (1990) 93
\item{57.} J. Honerkamp, Nucl. Phys {\bf B36} (1972) 130;\hfill \break
           S. Randjbar-Daemi, A. Salam and J.A. Strathdee, Int. J. Mod. Phys
           {\bf A2} (1987) 667;\hfill \break
           A. Tseytlin, Phys. Lett. {\bf 223B} (1989) 165;
           \hfill \break
           O. Andreev, R. Metsaev and A. Tseytlin, Sov. J. Nucl. Phys
           {\bf 51(2)} (1990) 359 \hfill \break
            \vfill\eject\end